# Implications of Interactions between Steric Effects and Electrical Double Layer Overlapping Phenomena on Electro-Chemical Transport in Narrow Fluidic Confinements

Siddhartha Das and Suman Chakraborty[1]

*Department of Mechanical Engineering, IIT Kharagpur-721302, India*

[1]Corresponding author, e-mail: suman@mech.iitkgp.ernet.in

**Abstract**

In this paper, we study the non-trivial interactions between the ionic Steric effects and Electric Double Layer (EDL) overlap phenomenon on the resultant electro-chemical transport in narrow fluidic confinements. Through a comprehensive mathematical model, we demonstrate that more prominent Steric effects may result in greater magnitudes of the channel centerline potential. However, since the magnitude of the zeta potential also gets perpetually enhanced with Steric interactions, this phenomenon cannot by be considered by itself as a trivial interpreter of an augmentation in the extent of the effective EDL overlap. Further investigations in this regard, however, do reveal an intricate coupling between EDL overlap phenomenon and finite ionic size effects, so as to result in an effective enhancement in the extent of EDL overlap, far beyond what is predicted by classical electrochemical considerations. Insights are also provided on the possible implications of the intricate interactions of the underlying physico-chemical mechanisms on the design of future-generation nanofluidic devices.

# I. INTRODUCTION

Rapid advancements in miniaturized fabrication technologies have opened up huge possibilities of investigating the electro-chemical-hydrodynamics in narrow fluidic confinements and exploiting the same in developing nano-scale functional devices that find wide applications in virtually every sphere of modern-day human needs.[1-8] A comprehensive understanding of the involved physics in many such intricate devices often necessitates the study of a double layer of electrical charge (popularly termed as Electric Double layer or EDL[9]) that screens the bare charge of the channel walls. The problem of decay of the electrical potential within the EDL is not only of immense contemporary relevance, but also is of classical importance, owing to its fundamental significance in electrochemistry and surface sciences.[9-12]

In its most simplified form, the electrostatic problem of EDL charge distribution in a fluidic confinement is typically solved under various simplifying assumptions in addition to the consideration of a mean field approximation, such as non-overlapping EDLs (i.e., the characteristic EDL thickness, $\lambda$, which is typically of the order of few nanometers, is smaller than the channel hydraulic radius) or the assumption that the ions within the EDL can be treated as point charges (i.e., with no ionic size effect).[9] Such simplifications have been found to work fairly well for many practical problems. However, of late, with the growing need of studying systems with large relative EDL thicknesses and/ or large zeta potentials, such simplified assumptions have been found to be grossly violated in many cases. Analysis of transport phenomena in such devices, in effect, is likely to demand a simultaneous consideration of overlapped EDL conditions and finite size effects of the ionic species. From fundamental thermodynamic considerations, consequences of these two effects may apparently differ significantly. For the case with EDL overlap, the energy picture is still the balance between the thermal energy and the electrostatic energy of the ions, despite the existence of a non-zero centerline electrostatic potential due to EDL effects protruding into the bulk. However, for the case where the finite size of the EDL ions become important, an additional contribution due to size-induced mixing entropy appears and must be accounted for in the calculation.

A detailed review of the literature on the effects of overlapping EDLs (within continuum limits) on the consequent transport processes can be found in the recent paper by Baldessari and Santiago.[13] As discussed in the above work, the most significant difficulty in handling the calculation with overlapped EDLs is the correct specification of the condition (ionic concentrations and EDL potential) at the channel centerline, which, unlike for the case with non-overlapped EDL conditions, are not known *apriori*. Most of the models[14-24] referred to in the paper by Baldessari and Santiago[13] tackle this problem by solving the EDL potential under the assumption of net charge electroneutrality of an isolated channel. Such modeling strategies invoke the considerations of reaction equilibrium appealing to the detailed surface chemistry at the channel-fluid interface, and relating the same with the overall charge neutrality and mass conservation constraints. As an alternative approach, Baldesaari and Santiago[13] considered the channel to be connected by infinite ionic reservoirs, allowing the specification of the channel centerline conditions in terms of the known conditions at the infinite reservoirs (or wells). Using such considerations, they presented detailed calculations on nanochannel transport under overlapped EDL conditions. Other related recent studies on the applications and consequences of overlapping EDLs include the investigation of double layer overlap-induced change in pressure gradient and electrical conductivity in microchannels,[25] alteration in energy conversion in nanofluidic channels,[26] alteration of slip behavior in pressure-driven electrokinetic microchannel transport,[27] enhanced surface interactions between like-charged polyelectrolyte gels,[28] augmented ion-current rectification through nanopores,[29,30] enhancement of nanochannel electrokinetic pumping efficiency,[31] alterations in AC electrokinetics,[32,33] correct estimation of the induced pressure gradients due to entrance and exit effects in electroosmotic flows,[34] etc.

It is also important to mention in the present context that notwithstanding the extent of the EDL interactions, fundamental transport within the same has been investigated in the literature with varying simplifying assumptions. One of the traditional models commonly invoked in this regard is the Gouy-Chapman (GC) model, which treats the solvent as a continuum and the ionic species as point charges. Following this model, the potential distribution within the EDL is described by the celebrated Poisson Boltzmann (PB) equation. Such descriptions, however, have been found to be adequate only when the surface charge densities and electrolyte concentrations are sufficiently low. For higher ionic concentrations and/or higher surface charge densities, discrete natures of the ionic species turn out to be progressively more consequential. This may be explained as follows, following Kilic et al.[35] Considering characteristic length scale of each ionic species to be *a*, the corresponding maximum number density of ions turns out to be $\frac{1}{a^3}$. Following the Boltzmann distribution of ionic charges (with $k_B$ as the Boltzmann constant, $T$ as the absolute temperature, $e$ as the protonic charge, and $z$ as the ionic valency), the critical potential at which this number density is reached is given by $\psi_c = \frac{k_B T}{ze} \ln\left(\frac{2}{\upsilon}\right)$, where $\upsilon = 2n_\infty a^3$ is the so-called steric factor, which is physically an indicator of the bulk volume fraction of ions. This, in turn, implicates the order of magnitude of the critical potential as $\frac{k_B T}{ze}$, beyond which the PB description may yield an unphysical picture, even for moderately dilute solutions. As a consequence, the GC model predicts impractically large counter-ionic concentrations adjacent to highly charged substrates. In an effort to overcome such constraints, Bikerman[36,37] derived a modified form of the PB equation, accounting for the repulsions between the finite-sized ionic species. Wicke and Eigen[38-40] were amongst the others who attempted to modify the dilute solution treatment of the PB equation by incorporating the excluded volume effect of ions, which introduces an additional molecular length scale $a_0$ in the formulation. Their theory was further refined by contributions from Iglic and coworkers[41-44] and Borukhov, Andelman and Orland.[45-47] The principal idea of this theory is to incorporate the size effect of ions by including the entropy contribution of the finite (equal) sized ions in the free energy and then minimizing this free energy to obtain the modified PB equation. All the derivations were made under the classical mean field approximation. There has also been significant number of efforts to study the finite ion size effect in EDL potential estimation beyond mean field approximation, i.e., by invoking non-continuum approaches like molecular dynamics simulation or Monte Carlo simulations.[48-57] For the present study, however, we only concentrate on the mean field approach, which neglects the discrete, many body interactions and as has been pointed out by Kilic et al.,[35] such an approach can indeed lead to satisfactory accounting of the effect of finite ion sizes for a large number of practical situations. There are also instances of recent investigations that invoke this mean field approach to estimate the influence of Steric effect on different classical nanofluidic electrokinetic phenomena such as generation of nanofluidic streaming current,[58] EDL-induced modification of wetting tension of a charged electrolytic drop,[59] double layer polarization of a non-conducting spherical particle in a time varying electric field,[60] electrophoresis of a colloidal sphere,[61] AC electroosmosis in dilute electrolytes[62] etc.

From the above discussions, it is clear that there have been large number investigations that account for either EDL overlap or finite ion size effect in EDL potential estimation. In reality, however, these effects may act synergistically in tandem and interact with each other non-trivially, especially for confinements with large relative EDL thicknesses under large surface charge density conditions. However, to the best of our knowledge, there exists no study in the literature that calculates the EDL potential by considering the EDL overlap and the Steric effect simultaneously in a self-consistent mathematical framework. Such a model, if framed, will be the most generalized mean field representation of the EDL potential distribution hitherto undressed in the literature. In this paper we attempt to frame and solve such a model towards obtaining the EDL potential distribution. The differential equations governing the EDL potential and the coion and counterion distributions are obtained by minimizing the free energy, which has the additional entropy contribution owing to the finite ion sizes. These equations are then integrated in presence of the appropriate boundary conditions that account for the consideration of overlapped EDLs. Numerical solutions are obtained for chemical reaction equilibrium condition at the wall (atypical of bare Silica wall nanochannels). Results from the present investigation essentially implicate that the Steric effects act so as enhance the effective extent of EDL overlap based on the relevant parametric space, as consequence of intricate interactions between modified ionic conditions as attributable to finite ion size effects, and the involved electro-chemical transport as accountable to overlapped EDL conditions.

**II. THEORY**

We consider a narrow fluidic confinement connected between two large ionic reservoirs (see figure 1). In these two reservoirs, the concentration of the cations and the anions are constants (at $n_\infty$). The channel walls are assumed to be located at $y = \pm H$ (see figure 1). The channel is considered to be narrow enough so that the EDLs formed at the two opposite walls can significantly overlap. In this paper we frame and solve the modified PB equation for the resulting EDL potential distribution in the channel with highly overlapping EDLs, additionally accounting for the effect of the finite size of the ionic species as well.

The fact of EDL overlap essentially implicates that the counterion and coion concentrations at the channel centerline are not identical. Here we express those as $n_+ = n_{+,c}$, $n_- = n_{-,c}$ and the resulting potential as $\psi = \psi_c$. Following the approach proposed by Baldessari,[13] one can relate $n_{\pm,c}$ with $\psi_c$ as $n_{\pm,c} = n_\infty \exp\left(\mp \frac{ez\psi_c}{k_B T}\right)$ (the necessary physical conditions that are implied in this assumption are detailed by Baldessari[13] and are not repeated here for the sake of brevity). Based on this fundamental condition, we now proceed to derive the modified Poisson-Boltzmann equation, by invoking the standard free energy formulation, as outlined elsewhere.[35,47]

We consider a symmetric z:z electrolyte for which the free energy is assumed to be of the form

$$F = U - TS \tag{1}$$

where

$$U = \int d\mathbf{r}\left(-\frac{\varepsilon}{2}|\nabla\psi|^2 + zen_+\psi - zen_-\psi\right) \tag{2}$$

and

$$-TS = -\frac{k_B T}{a^3}\int d\mathbf{r}\left\{n_+ a^3 \ln(n_+ a^3) + n_- a^3 \ln(n_- a^3) + (1 - n_+ a^3 - n_- a^3)\ln(1 - n_+ a^3 - n_- a^3)\right\} \tag{3}$$

In eq. (2), the first term within the parenthesis in the right hand side represents the self free energy density of the EDL field and the remaining two terms are the free energy densities due to the electrostatic interaction of the ions with the EDL field. In eq. (3), all the terms within the parenthesis in the right hand side represent the free energy density due to entropic mixing of the ions in the bulk. Finally the parameter "$a$" is the effective size of the ions and the solvent molecules. The ionic concentration distributions are derived assuming the net electrochemical potential of each type of ions (which we denote as $\mu_\pm = \frac{\partial F}{\partial n_\pm}$) is constant in the system. We express

$$\mu_\pm = \frac{\partial F}{\partial n_\pm} = \int d\mathbf{r}\left[\pm ez\psi + k_B T \ln\left(\frac{n_\pm a^3}{1 - n_+ a^3 - n_- a^3}\right)\right] \tag{4}$$

As the differential $d\mathbf{r}$ represents any arbitrary volume in the system, constancy of $\mu_\pm$ will mean

$$\pm ez\psi + k_B T \ln\left(\frac{n_\pm a^3}{1 - n_+ a^3 - n_- a^3}\right) = \text{constant} \tag{5}$$

Thus taking differential on both sides of eq. (5), one gets:

$$\frac{dn_\pm}{n_\pm} + \frac{a^3(dn_+ + dn_-)}{1 - a^3(n_+ + n_-)} = \mp \frac{ez}{k_B T} d\psi \tag{6}$$

Writing $1 - a^3(n_+ + n_-) = p$, eq. (6) can be rewritten as:

$$\frac{dn_\pm}{n_\pm} - \frac{dp}{p} = \mp \frac{ez}{k_B T} d\psi \tag{7}$$

Integrating eq. (7), we get

$$\ln\left(\frac{n_\pm}{p}\right) = \mp \frac{ze\psi}{k_B T} + K_\pm \tag{8}$$

where $K_\pm$ is the constant of integration. To evaluate $K_\pm$, we apply the condition that at the channel centerline we have $n_\pm = n_{\pm,c} = n_\infty \exp\left(\mp \frac{ez\psi_c}{k_B T}\right)$ and $\psi = \psi_c$. Noting that $p = 1 - (n_{+,c} + n_{-,c})a^3 = 1 - 2n_\infty a^3 \cosh\left(\frac{ez\psi_c}{k_B T}\right) = 1 - v\cosh\left(\frac{ez\psi_c}{k_B T}\right)$, where $v = 2n_\infty a^3$ is the Steric or size factor, we obtain:

$$K_\pm = \ln\left(\frac{n_{\pm,c}}{1 - v\cosh\left(\frac{ez\psi_c}{k_B T}\right)}\right) \pm \frac{ez\psi_c}{k_B T} \tag{9}$$

Thus, the final ionic distribution equation reads:

$$\frac{n_+}{1 - a^3(n_+ + n_-)} = \frac{n_\infty}{1 - v\cosh\left(\frac{ez\psi_c}{k_B T}\right)} \exp\left(-\frac{ez\psi}{k_B T}\right) \tag{10a}$$

and

$$\frac{n_-}{1 - a^3(n_+ + n_-)} = \frac{n_\infty}{1 - v\cosh\left(\frac{ez\psi_c}{k_B T}\right)} \exp\left(\frac{ez\psi}{k_B T}\right) \tag{10b}$$

Eq. (10a and 10b), as it appears is not explicit in either $n_+$ or $n_-$. To obtain that explicit form, we first divide eq. (10a) by eq. (10b) to obtain

$$n_+ = n_- \exp\left(-\frac{2ez\psi}{k_B T}\right) \tag{11}$$

Using eq. (11) in (10b), we obtain (after some trivial algebra):

$$n_- = \frac{n_\infty \exp(\alpha)}{1 + \nu\left[\cosh(\alpha) - \cosh(\alpha_c)\right]} \quad \text{(where } \alpha = \frac{ez\psi}{k_B T} \text{ and } \alpha_c = \frac{ez\psi_c}{k_B T}\text{)} \tag{12}$$

Using eq. (12) in (11) we also get:

$$n_+ = \frac{n_\infty \exp(-\alpha)}{1 + \nu\left[\cosh(\alpha) - \cosh(\alpha_c)\right]} \tag{13}$$

To obtain the equation governing the potential distribution, we minimize the free energy expression with respect to the variable $\psi$, so as to obtain:

$$\frac{\partial F}{\partial \psi} = 0 = \int d\mathbf{r}\left[-\frac{\partial}{\partial \psi}\left\{\frac{\varepsilon}{2}|\nabla \psi|^2\right\} + zen_+ - zen_-\right] \tag{14}$$

As the differential $d\mathbf{r}$ represents any arbitrary volume in the system, eq. (14) will mean (using eqs. (12) and (13)):

$$-\frac{\partial}{\partial \psi}\left\{\frac{\varepsilon}{2}|\nabla \psi|^2\right\} + zen_+ - zen_- = 0$$

$$\Rightarrow \nabla^2 \psi = -\frac{zen_\infty}{\varepsilon} \frac{\exp(-\alpha) - \exp(\alpha)}{1 + \nu\left[\cosh(\alpha) - \cosh(\alpha_c)\right]} = \frac{zen_\infty}{\varepsilon} \frac{2\sinh\left(\frac{ez\psi}{k_B T}\right)}{1 + \nu\left[\cosh\left(\frac{ez\psi}{k_B T}\right) - \cosh\left(\frac{ez\psi_c}{k_B T}\right)\right]} \tag{15}$$

Assuming one dimensional variation of $\psi$, we can write:

$$\frac{d^2\psi}{dy^2} = \frac{zen_\infty}{\varepsilon} \frac{2\sinh\left(\frac{ez\psi}{k_B T}\right)}{1 + \nu\left[\cosh\left(\frac{ez\psi}{k_B T}\right) - \cosh\left(\frac{ez\psi_c}{k_B T}\right)\right]} \tag{16}$$

Multiplying both sides of eq. (16) with $\frac{d\psi}{dy}$, we can write:

$$\frac{d\psi}{dy}d\left(\frac{d\psi}{dy}\right) = \frac{zen_\infty}{\varepsilon}\frac{2\sinh\left(\frac{ez\psi}{k_BT}\right)}{1+\nu\left[\cosh\left(\frac{ez\psi}{k_BT}\right)-\cosh\left(\frac{ez\psi_c}{k_BT}\right)\right]}d\psi \tag{17}$$

Writing $q = 1+\nu\left[\cosh\left(\frac{ez\psi}{k_BT}\right)-\cosh\left(\frac{ez\psi_c}{k_BT}\right)\right]$, we get:

$$\frac{d\psi}{dy}d\left(\frac{d\psi}{dy}\right) = \frac{zen_\infty}{\varepsilon}\left(\frac{k_BT}{ez\nu}\right)\frac{dq}{q} \tag{18}$$

On integrating eq. (18), we get:

$$\left(\frac{d\psi}{dy}\right)^2 = 4\left(\frac{n_\infty k_BT}{\varepsilon\nu}\right)ln\left\{1+\nu\left[\cosh\left(\frac{ez\psi}{k_BT}\right)-\cosh\left(\frac{ez\psi_c}{k_BT}\right)\right]\right\} + C \tag{19}$$

At the channel centerline $\frac{d\psi}{dy} = 0$ and $\psi = \psi_c$, which gives $C = 0$. Hence, we have:

$$\left(\frac{d\psi}{dy}\right) = \pm 2\sqrt{\frac{n_\infty k_BT}{\varepsilon\nu}}\sqrt{ln\left\{1+\nu\left[\cosh\left(\frac{ez\psi}{k_BT}\right)-\cosh\left(\frac{ez\psi_c}{k_BT}\right)\right]\right\}} \tag{20}$$

As a verification of the correctness of this expression, we consider the case where there is no EDL overlap, i.e., $\psi_c = 0$ so that eq. (20) can be expressed as (using $\cosh\left(\frac{ez\psi}{k_BT}\right) - 1 = 2\sinh^2\left(\frac{ez\psi}{2k_BT}\right)$ and $\lambda_D = \sqrt{\frac{\varepsilon k_BT}{2e^2z^2n_\infty}}$):

$$\left(\frac{d\psi}{dy}\right) = \pm 2\sqrt{\frac{n_\infty k_BT}{\varepsilon\nu}}\sqrt{ln\left\{1+\nu\left[\cosh\left(\frac{ez\psi}{k_BT}\right)-1\right]\right\}} = \pm 2\frac{ezn_\infty\lambda_D}{\varepsilon}\sqrt{\frac{2}{\nu}ln\left\{1+2\nu\sinh^2\left(\frac{ez\psi}{k_BT}\right)\right\}} \tag{21}$$

The expression in eq. (21) is identical to that derived by Kilic et al.,[35] under non-overlapped EDL conditions. On the other hand, the case without Steric effects but with EDL overlap can be recovered by setting $\nu \to 0$ in Eq. (20), so that one obtains

$$\left(\frac{d\psi}{dy}\right) = \pm 2\sqrt{\frac{n_\infty k_BT}{\varepsilon}}\sqrt{\left[\cosh\left(\frac{ez\psi}{k_BT}\right)-\cosh\left(\frac{ez\psi_c}{k_BT}\right)\right]}$$, using the mathematical condition

$\left(\frac{ln(1+x)}{x}\right)_{lim\,x\to 0} = 1$.

It is evident that the final solution of the EDL potential $\psi$ can only be obtained with the correct specification of the condition at the channel walls, as well as the specification of the potential at the channel centerline. To achieve this purpose, we consider here an illustrative example that the channel walls are made of bare Silica so that the zeta potential is determined from the equilibrium of the chemical reaction between the bare Silica, the hydrogen ions, and the added cations. Hence, the zeta potential depends on the bulk ionic concentration, $n_\infty$, and the buffer pH. Thus one can write, after accounting for the appropriate chemical reaction at the channel wall (for details of the reaction description one may refer to Behrens and Grier[63]), the interrelationship between the wall zeta potential ($\zeta$) and the bare wall charge density ($\sigma$) as:

$$\zeta = \frac{k_B T}{e} \ln \frac{-\sigma}{e\Gamma + \sigma} - \frac{k_B T}{e}(pH_0 - pK_a)\ln 10 - \frac{\sigma}{C_{St}} \quad (22)$$

where $\Gamma$ is the fraction of dissociated chargeable sites, $pH_0$ is the value of the pH of the solution in the reservoirs, $K_a$ is the dissociation constant of the silica-water interface and $C_{St}$ is the capacitance of the Stern layer. Also we can apply eq. (20) at *the channel walls* (where $\sigma = \pm \varepsilon \frac{d\psi}{dy}$) so that one can write (for negative zeta potential):

$$\sigma = -2\sqrt{\frac{n_\infty \varepsilon k_B T}{\nu}} \sqrt{\ln\left\{1 + \nu\left[\cosh\left(\frac{ez\zeta}{k_B T}\right) - \cosh\left(\frac{ez\psi_c}{k_B T}\right)\right]\right\}} \quad (23)$$

To self-consistently obtain $\zeta$, $\psi_c$ and $\sigma$ we approach as follows:

We first assume a zeta potential ($\zeta$) which is used to obtain $\psi_c$ (for a given set of parameters) by numerically integrating eq. (20) under the condition that at the centerline $\psi = \psi_c$. With this $\zeta$ and $\psi_c$, we first obtain $\sigma$ (from eq. 23), which is used in eq. (22) to obtain a new value of $\zeta$. This iteration is continued till all the variables $\zeta$, $\psi_c$ and $\sigma$ cease to change.

**III. RESULTS AND DISCUSSIONS**

For illustrating the significant implications of the present model, we choose the following parameters: $\Gamma = 5.0$ nm$^{-2}$, $pK_a = 7.5$ and $C_{St} = 0.3$ F/m$^2$. It is important to mention in this context that we have also obtained results using various other values of these parameters, within practical limits. Those results exhibit identical physical trends as compared to the ones presented in this work, and hence are not reported here for the sake of brevity.

Figures 2a, b depict variations in the zeta potential and centerline potential respectively, as a function of the Steric factor, ν, and the $\frac{\lambda}{H}$ ratio. It can be observed from these figures that for a given value of ν, the magnitude of ζ increases with pH and λ/H. This matches well with the experimentally reported qualitative dependences of ζ on pH and bulk ionic concentration (which dictates the Debye length, λ) for the case with no Steric effect and no EDL overlap.[64] Regarding the implications of the Steric effect, it may be noted that magnitudes of both ζ and $\psi_c$ are found to increase with increments in the Steric factor, ν as evident from figure 2. Augmentation in the magnitude of zeta potential with the Steric effect can be interpreted as a consequence of lesser concentration of counterions in the vicinity of the wall (and hence lesser screening of the wall charge) owing to steric hindrance. On the contrary, it is difficult to pinpoint whether the increase in the channel centerline potential is merely a sole artifact of the corresponding increase in the magnitude of the zeta potential, or is also contributed by the unique electro-physical condition induced due to the Steric effects in presence of overlapping EDLs. Before attempting to obtain a more plausible answer to this issue, we try to address whether the combined consequence of the Steric effect and overlapping EDLs on the zeta potential and the channel centerline potential is effectively a mere superposition of these two individual effects. In order to assess the underlying consequences, we introduce a parameter $\zeta_s$, which is the sum of the zeta potential obtained separately for overlapped EDL conditions with no steric effects ($\zeta|_{\lambda/H,\nu\rightarrow 0}$) and the zeta potential obtained separately in presence of steric effects but within thin EDL limits ($\zeta|_{\lambda/H\ll 1,\nu}$), so that $\zeta_s = \zeta|_{\lambda/H,\nu\rightarrow 0} + \zeta|_{\lambda/H\ll 1,\nu}$. Analogously, we introduce a parameter $\psi_{c,s}$ which is the sum of the centerline potential obtained separately for overlapped EDL conditions with no Steric effects ($\psi_c|_{\lambda/H,\nu\rightarrow 0}$) and the centerline potential obtained separately in presence of Steric effects but within thin EDL limits ($\psi_c|_{\lambda/H\ll 1,\nu}$), so that $\psi_{c,s} = \psi_c|_{\lambda/H,\nu\rightarrow 0} + \psi_c|_{\lambda/H\ll 1,\nu}$. In figure 3a, we compare ζ with $\zeta_s$, whereas in figure 3b we compare $\psi_c$ with $\psi_{c,s}$. From these figures it is clearly revealed that the combined consequence of Steric

effect and EDL overlap evokes a physical situation which cannot be trivially interpreted as due to their superposed influences. In presence of the reaction equilibrium at the channel walls, both larger ionic sizes and thicker EDLs implicate weaker screening of the bare wall charge and hence an augmented magnitude of $\zeta$ potential. Thus, fundamentally, the location at which the screening counterions are placed relative to the charged wall turns out to be the influencing factor that dictates the magnitude of the zeta potential. For smaller Steric factors, it is primarily the span of the EDL that governs this relative location of the counterions, and consequently zeta potential is mostly the representation of the EDL effect with the effect of the Steric influence being substantially overemphasized. Such masking of the finite ion size effect does not occur when the Steric effect and the overlapped EDL effects are separately considered, and hence for smaller ion sizes $|\zeta| < |\zeta_s|$, with the difference getting lowered for thicker EDLs. However, for larger ionic sizes an altogether different physical effect becomes relevant in dictating the ratio $\zeta/\zeta_s$. For larger ionic sizes, their influences in dictating relative location of the counterions (from the wall) can be as important as that of the EDL thickness. For such cases, estimation of $\zeta|_{\lambda/H \ll 1, \nu}$ will mean that the ions are forced into a negligibly small EDL thickness. Such positional constraint (or barrier) leads to a large loss of their mixing entropy (an analogy may be drawn with to the loss of entropy encountered by a polymer chain when forced through a narrow confinement owing to the loss of conformations [65]), as encountered during the calculation of $\zeta|_{\lambda/H \ll 1, \nu}$, as apart of the calculation for $\zeta_s$. Hence, for such cases, $|\zeta| > |\zeta_s|$.

Regarding the effect of pH, it can be noted that higher pH leads to greater ionization of the channel wall. Responding to that, larger number of counterions accumulates near the channel wall. This invariably leads to a more pronounced mixing effect, particularly for the case in which there is no additional positional constraint on the counterions (pertinent to the situation in which the Steric effect and the EDL overlap effects are simultaneously considered). Thus, for any given combination of ν and λ/H, $\zeta/\zeta_s$ always has a higher value for a greater pH value.

Unlike the zeta potential, the channel centerline potential always exhibits an enhanced value as compared to the prediction made by superposition of the individual effects (see fig. 3b). This is due to the fact that irrespective of the Steric effect, for very thin EDL, $\psi_c \to 0$. This clearly establishes that the consequence of the interactions of the effects of finite ion sizes and the overlapping EDLs are distinctly different for the zeta potential and the channel centerline potential. Alteration of the zeta potential is a combined consequence of the surface chemistry and the non-trivialities in the electrochemical interactions as introduced by the Steric effect. On the other hand, deviation of channel centerline potential from zero occurs only when there is EDL overlap and for such cases the Steric effect may amplify the effect of extent of this overlap to a considerable limit.

To provide a more concrete justification of the Steric effect induced enhancement in the extent of EDL overlap, we plot the transverse variation of the EDL potential field, made dimensionless with the corresponding zeta potential (see figs. 4a-d). Apart from representing the exact behavior of the EDL potential under hitherto unaddressed physical conditions, these plots also provide sufficient qualitative evidence that the channel centerline potential (for any value of buffer pH or relative EDL thickness) shows an enhancement that is beyond a value governed solely by the corresponding increase in the zeta potential This is evidenced by a continuously enhanced value of $\psi_c/\zeta$ for larger Steric factor $\nu$. Larger Steric factor, beyond increasing the wall zeta potential, signifies larger effective size of the counterions, enforcing those to remain excluded from locations very close to the wall. Such distinct positional characteristics of the counterions essentially implicates that the screening effect of the counterions is significantly lowered, thereby forcing a much weaker decay of the EDL potential with distance from the wall, which naturally signifies a more enhanced extent of the EDL overlap.

With a physical basis of the Steric effect induced enhancement of EDL overlap established as above, it may further be imperative to quantify this enhancement in terms of the pertinent electrochemical parameters. For that purpose, we obtain the variation of both zeta potential as well as the channel centerline potential with respect to a reference (Steric effect independent) zeta potential. This reference zeta potential is defined as:

$$\zeta_0 = \frac{k_B T}{e} ln \frac{-\sigma_0}{e\Gamma + \sigma_0} - \frac{k_B T}{e} (pH_0 - pK) ln 10 - \frac{\sigma_0}{C_{St}} \quad (23)$$

where $\sigma_0$ is the Steric factor independent charge density defined as:

$$\sigma_0 = -2\sqrt{n_\infty \varepsilon k_B T} \qquad (24)$$

If it is observed that such relative variation of the channel centerline potential with increase in the Steric factor is more than that corresponding to the relative variation of the zeta potential, one may infer that the Steric effect indeed enhances the extent of EDL overlap. To quantify the above, we define two dimensionless ratios expressed as:

$$P_\zeta = \frac{(\zeta/\zeta_0) - (\zeta/\zeta_0)_{v \to 0}}{(\zeta/\zeta_0)_{v \to 0}} \qquad (25a)$$

and

$$P_\psi = \frac{(\psi_c/\zeta_0) - (\psi_c/\zeta_0)_{v \to 0}}{(\psi_c/\zeta_0)_{v \to 0}} \qquad (25b)$$

From eqs. (25a) and (25b), it is clear that these two ratios indicate the relative variation of the zeta potential and the channel centerline potential with increase in the Steric effect. In fig. (5), we plot the variation of the ratio $R = \frac{P_\psi - P_\zeta}{P_\zeta}$. It is clearly demonstrated that for all combinations of the Steric factor, EDL thickness and buffer pH, $P_\psi > P_\zeta$ and the increase is more prominent at a higher value of the Steric factor. For either of the studied pH values, the Steric effect induced EDL overlap enhancement is more prominent for that cases with originally smaller values of EDL overlap. This can be argued from the fact that the Steric effect has relatively lesser consequence in a system that itself originally has extremely thick EDLs. From Fig. 5 it is conclusively established that the ratio $R$ may act as a pertinent dimensionless parameter that quantifies the extent of EDL overlap due to Steric effects, as a function of other relevant system parameters, in a physically consistent manner.

## IV. CONCUSIONS

In this paper, we investigate the effects of finite ionic sizes on the EDL potential distribution in narrow fluidic confinements under overlapped EDL conditions. We establish that the Steric effect can significantly enhance both the zeta potential as well as the channel centerline potential. Accordingly, the Steric effect can lead to a significant enhancement in the extent of the effective EDL overlap, well beyond that predicted from the standard considerations of EDL interactions in which the ions are taken as point charges. We also establish that the combined interaction of the Steric effect and overlapping EDLs leads to conditions, which, depending on the value of the Steric factor, triggers a physical behavior that is substantially different from that obtained by mere superposition of the individual consequences of EDL overlap without Steric effects and Steric effects without EDL overlap. Proceeding further, we quantify the enhancement in effective the EDL overlap as a function of the ionic Steric factor, for different relative Debye lengths.

Our findings can have immense relevance to several important open questions in the area of nano-electrokinetics. There has been continuous endeavor to design and fabricate nanochannels with significant extent of EDL overlap, which invariably optimizes several nano-electrokinetic outputs such as streaming current,[66] energy conversion efficiency,[67] faster separation[68] etc. But the desired requirement of relative EDL thickness needs to be often sacrificed on the basis of constraints such as precision on nanochannel fabrication, system capability of handling of ionic buffers etc. In this light, the present study can be extremely useful in the sense that with an optimal combination of the EDL overlap and the finite Steric factor effects, researchers can now access regimes of effectively augmented overlapped EDL conditions, even within the practically achievable conditions of nanofabrication and buffer handling, by exploiting Steric interactions to a favourable extent.

**Figures**

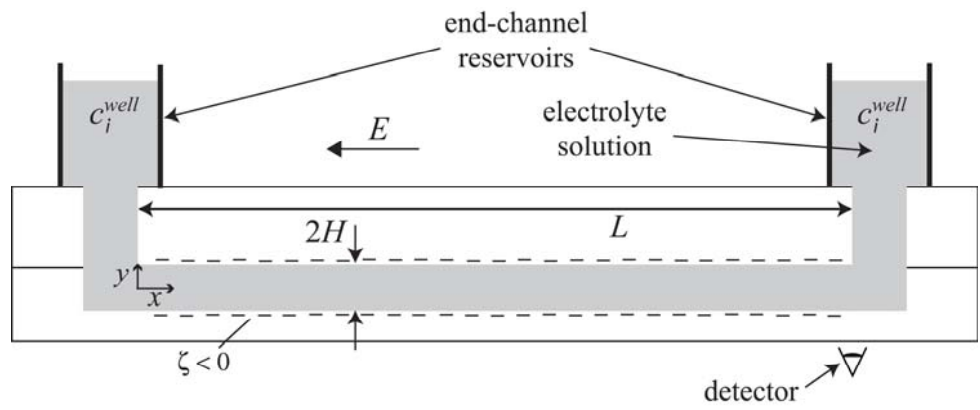

**Figure 1:** Schematic of the system

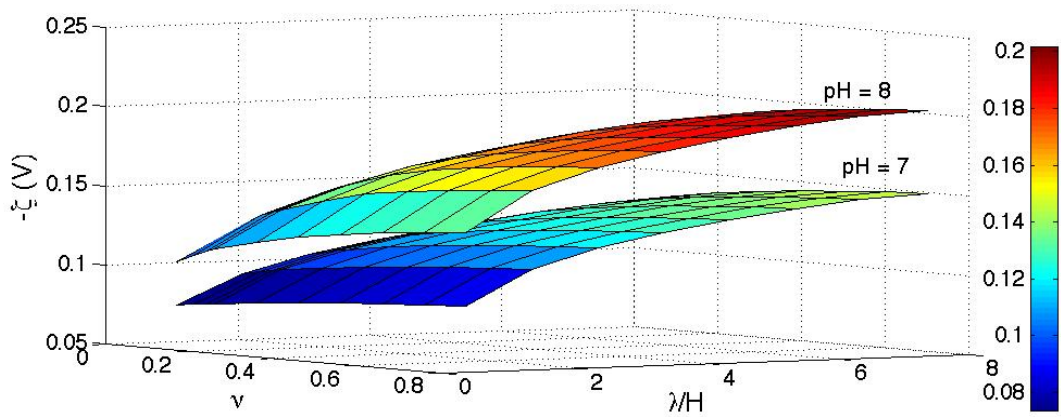

(a)

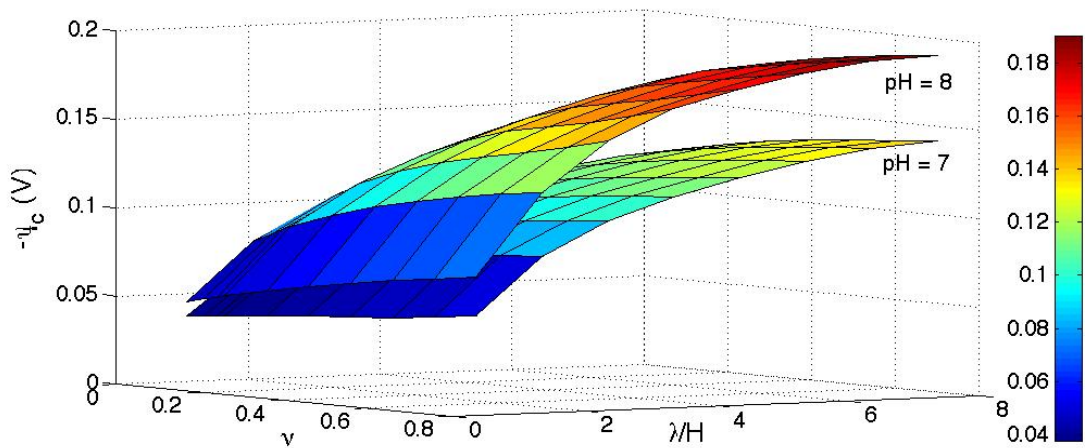

(b)

**Figure 2.** **(a)** Variation of zeta potential ($\zeta$) with Steric factor ($\nu$) and the relative EDL thickness ($\lambda/H$) for different values of buffer pH, **(b)** Variation of channel centerline potential ($\psi_c$) with Steric factor ($\nu$) and the relative EDL thickness ($\lambda/H$) for different values of buffer pH.

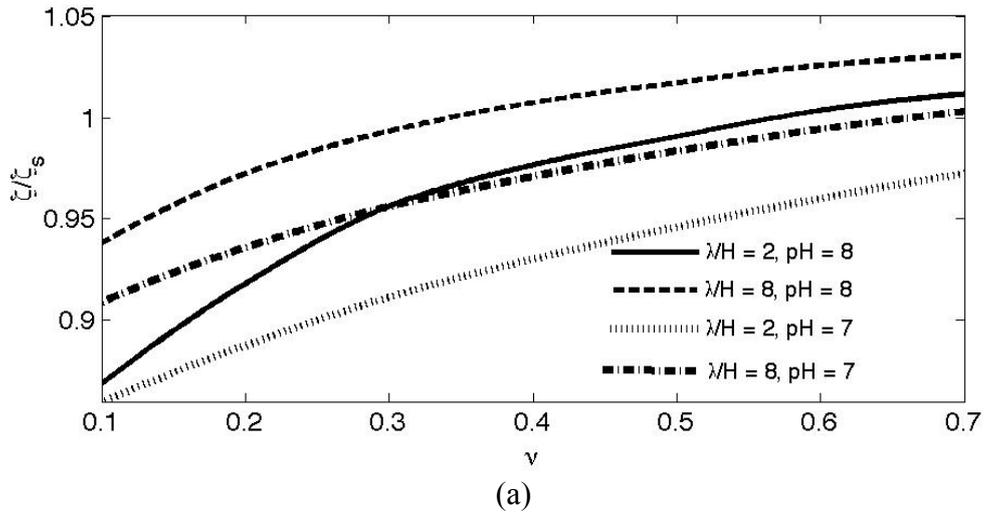

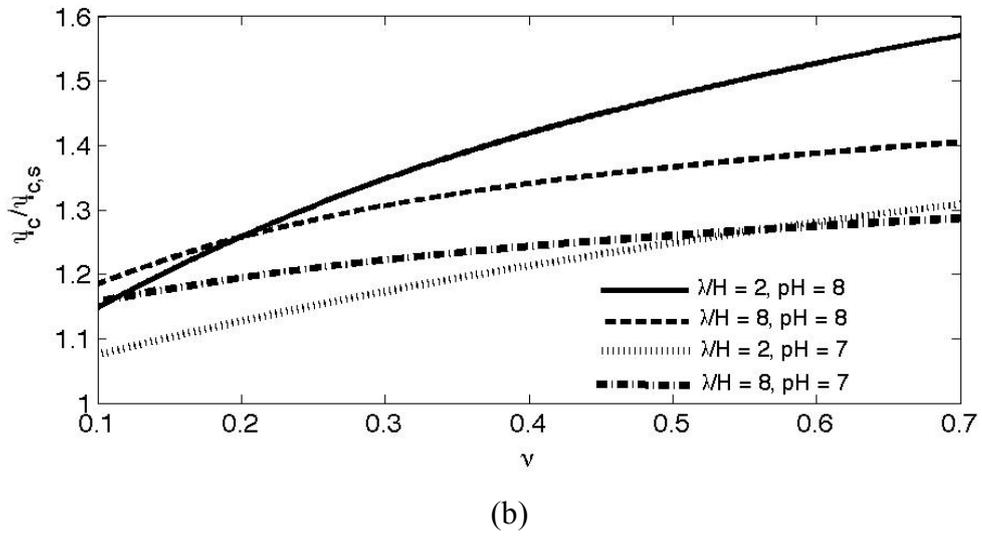

**Figure 3.** **(a)** Variation of the ratio $\zeta/\zeta_s$ (see text for detailed description on $\zeta_s$) with Steric factor ($\nu$) for different values of relative EDL thickness ($\lambda/H$) and buffer pH. **(b)** Variation of the ratio $\psi_c/\psi_{c,s}$ (see text for detailed description on $\psi_{c,s}$) with Steric factor ($\nu$) for different values of relative EDL thickness ($\lambda/H$) and buffer pH.

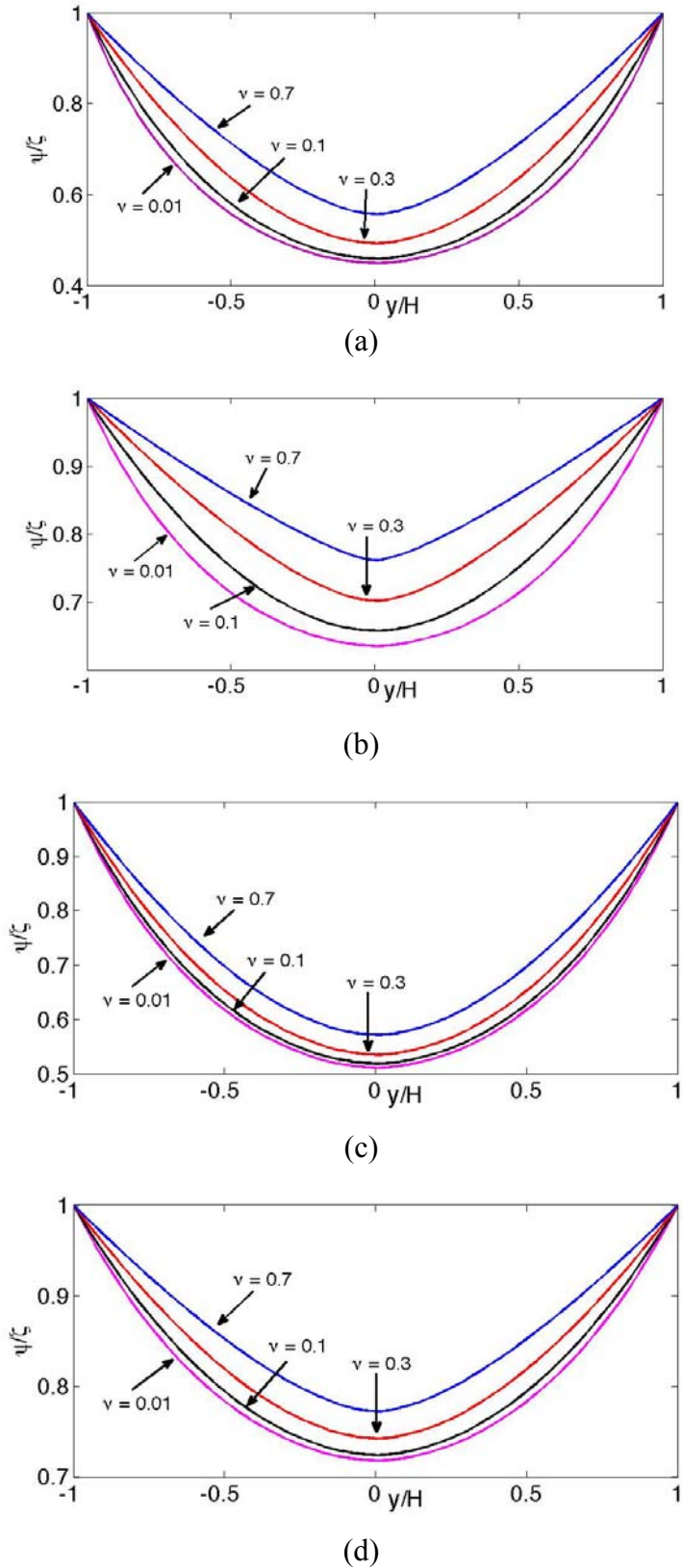

**Figure 4:** Transverse variation of EDL potential for different values of the Steric factor (ν) for the case with chemical equilibrium boundary condition for (a) $\lambda/H = 1$, pH = 8, (b) $\lambda/H = 2$, pH = 8 (c) $\lambda/H = 1$, pH = 7 and (d) $\lambda/H = 2$, pH = 7. For each of the plots, the zeta potential is that for the given combination of ν, $\lambda/H$ and pH.

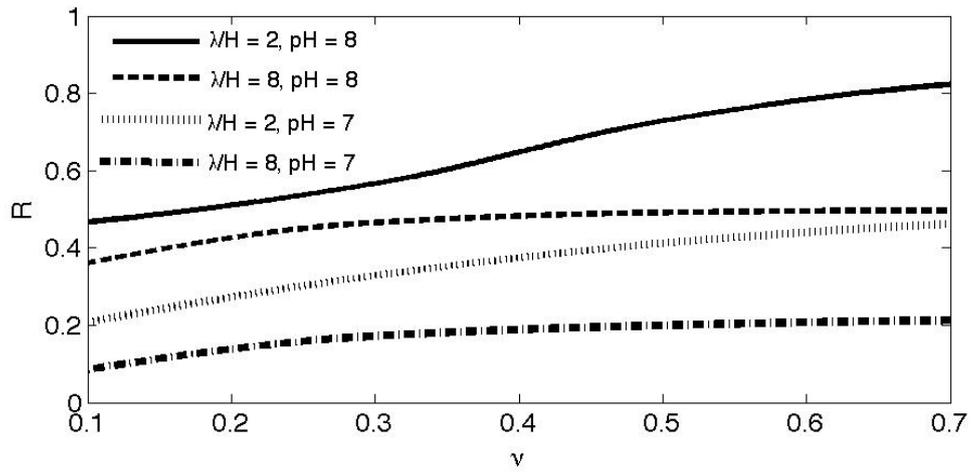

**Figure 5:** Variation of the ratio $R = \dfrac{P_\psi - P_\zeta}{P_\zeta}$ with the Steric factor $\nu$ for different values of the relative EDL thickness ($\lambda/H$) and buffer pH.